# No SQL, No Injection?

## Examining NoSQL Security

Aviv Ron
Cyber security Center of Excellence
IBM
Beer Sheba, Israel
rona@il.ibm.com

Alexandra Shulman-Peleg
Cyber security Center of Excellence
IBM,
Beer Sheba, Israel
shulmana@il.ibm.com

Emanuel Bronshtein
Application Security Research
IBM
Herzliya, Israel
emanuelb@il.ibm.com

*Abstract*—NoSQL data storage systems have become very popular due to their scalability and ease of use. This paper examines the maturity of security measures for NoSQL databases, addressing their new query and access mechanisms. For example the emergence of new query formats makes the old SQL injection techniques irrelevant, but are NoSQL databases immune to injection in general? The answer is NO. Here we present a few techniques for attacking NoSQL databases such as injections and CSRF. We analyze the source of these vulnerabilities and present methodologies to mitigate the attacks. We show that this new vibrant technological area lacks the security measures and awareness which have developed over the years in traditional RDBMS SQL systems.

## I. INTRODUCTION

Database security has been and will continue to be one of the more critical aspects of information security. Access to enterprise database grants an attacker a great control over the most critical data. For example, SQL injection attacks insert malicious code into the statements passed by the application to the database layer. This enables the attacker to do almost anything with the data including accessing unauthorized data as well as altering, deleting and inserting new data. Although the exploitation of SQL injection has been declining steadily over the years due to secure frameworks and improved awareness it has remained a high impact means to exploit system vulnerabilities. For example, it was shown that web applications receive 4 or more web attack campaigns per month and SQL injections are the most popular attacks on Retailers [1]. Lately NoSQL databases have emerged and are becoming more and more popular. Such databases for example are MongoDB [2], Redis [3], and Cassandra [4]. Some of these NoSQL databases use different query languages which make the traditional SQL injection techniques irrelevant. But does that mean NoSQL systems are immune to injections? Our study shows that while the security of the query language itself and the drivers has largely improved, there are still techniques for injecting malicious queries. In this paper we wish to raise the awareness of developers and information security owners to NoSQL security – focusing on the dangers and their mitigations. We present new injection techniques and discuss approaches for the mitigation of such attacks such as PHP array injection attack, MongoDB OR injection, arbitrary JavaScript injection and more.

## II. NOSQL

NoSQL (Not Only SQL) is a trending term in modern data stores. These are non-relational databases that rely on different storage mechanisms such as document store, key-value store, graph and more. The wide adoption of these databases is facilitated by the new requirements of modern large scale applications (e.g. Facebook, Amazon, Twitter) which need to distribute the data across a huge number of servers. These scalability requirements cannot be met by traditional relational databases which require that all operations of the same transaction are executed by a single database node [5][6]. According to accepted database popularity ranking three of the most common NoSQL databases (MongoDB, Cassandra and Redis) are ranked among the 10 most popular databases [10] and the popularity of NoSQL databases is constantly growing over the last years [11] . Like almost every new technology, NoSQL databases were lacking security when they first emerged [7] [8]. They suffered from lack of encryption, proper authentication and role management as well as fine grained authorization [9], Furthermore, they allowed dangerous network exposure and denial of service attacks [7]. Today the situation is better and popular databases introduced built-in protection mechanisms (e.g.,[23]). Yet, many best practices from traditional SQL databases are overlooked and the security of NoSQL deployments has not matured enough. In this paper we extend the observations of Okman et al [8] by providing detailed examples of NoSQL injection attacks. We describe CSRF vulnerabilities and discuss the actions needed to mitigate the risks of NoSQL attacks.

## III. JSON QUERIES AND DATA FORMATS

In the following sections we demonstrate how the popular JSON representation format allows new types of injection attacks. We illustrate this on the example of MongoDB, which is one of the most popular NoSQL databases [10]. MongoDB is a document-oriented database, which has been adopted for usage by multiple large vendors such as EBay, Foursquare, LinkedIn and others [13].

Queries and Data are represented in JSON format, which is better than SQL in terms of security because it is more "well defined", very simple to encode/decode and also has good native implementations in every programming language. Breaking the query structure as has been done in SQL injection is harder to do with a JSON structured query. A typical insert statement in MongoDB looks like:

```
db.books.insert({
    title: 'The Hobbit',
    author: 'J.R.R. Tolkien'
})
```

This inserts a new document into the books collection with a title and author field. A typical query looks like:

```
db.books.find({ title: 'The Hobbit' })
```

Queries can also include regular expressions, conditions, limit which fields get queried and more.

## IV. PHP ARRAY INJECTIONS

Let us examine an architecture depicted in Figure 1, where a web application is implemented with a PHP backend, which encodes the requests to the JSON format used to query the data store. Let's use an example of MongoDB to show an array injection vulnerability – an attack similar to SQL injection in its technique and results.

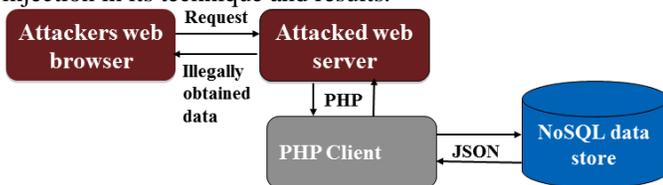

Figure 1: Architecture of a PHP web application

PHP encodes arrays to JSON natively. So for example the following array:

```
array('title' => 'The hobbit', 'author' => 'J.R.R. Tolkien');
```

would be encoded by PHP to the following json:

```
{"title": "The hobbit", "author": "J.R.R. Tolkien"}
```

Lets consider the following situation: A PHP application has a login mechanism where username and password are sent from the users browser via HTTP POST (the vulnerability is applicable also HTTP GET as well). A typical POST payload would look like:

username=tolkien&password=hobbit

And the backend PHP code to process it and query MongoDB for the user would look like:

```
db->logins->find(array("username"=>$_POST["username"],
"password"=>$_POST["password"]));
```

This makes perfect sense and is intuitively what the developer is likely to do, intending a query of:

```
db.logins.find({ username: 'tolkien', password: 'hobbit' })
```

But PHP has a built in mechanism for associative arrays which allows an attacker to send the following malicious payload:

username[$ne]=1&password[$ne]=1

PHP translates this input into:

```
array("username" => array("$ne" => 1), "password" =>
array("$ne" => 1));
```

Which is encoded into the mongo query:

```
db.logins.find({ username: { $ne: 1 }, password: { $ne: 1 } })
```

Since $ne is the not equals condition in MongoDB, this is querying all the entries in the logins collection where the username is not equal to 1 and the password is not equal to 1 which means this query will return all the users in the logins collection, in SQL terminology this is equivalent to:

```
SELECT * FROM logins WHERE username <> 1 AND password <> 1
```

In this scenario the vulnerability will lead to the attacker being able to log in to the application without a username and password. In other variants the vulnerability might lead to illegal data access or privileged actions performed by an unprivileged user. To mitigate this issue it is needed to cast the parameters received from the request to the proper type, in this case string.

## V. NOSQL OR INJECTION

One of the common reasons for a SQL injection vulnerability is building the query from string literals which include user input without using proper encoding. The JSON query structure makes it harder to achieve in modern data stores like MongoDB. Nevertheless it is still possible. Let us examine a login form which sends its username and password parameters via an HTTP POST to the backend which constructs the query by concatenating strings. For example the developer would do something like:

```
string query = "{ username: '" + post_username + "', password: '" + post_password + "' }"
```

With valid input (tolkien + hobbit) this would build the query:

```
{ username: 'tolkien', password: 'hobbit' }
```

But with malicious input this query can be turned to ignore the password and login into a user account without the password, here is an example for malicious input:

username=**tolkien', $or: [ {}, { 'a':'a**&password=**' ],  $comment:'successful MongoDB injection'**

This input will be constructed into the following query:

```
{ username: 'tolkien', $or: [ {}, { 'a': 'a', password: '' } ], $comment: 'successful MongoDB injection' }
```

This query will succeed as long as the username is correct. In SQL terminology this query is similar to:

```
SELECT * FROM logins WHERE username = 'tolkien' AND (TRUE OR ('a'='a' AND password = '')) #successful MongoDB injection
```

That is, the password becomes a redundant part of the query since an empty query {} is always true and the comment in the end does not affect the query. How did this happen? Let's examine the constructed query again and color the user input in bold red and the rest in black:

```
{ username: 'tolkien', $or: [ {}, { 'a': 'a', password: '' } ], $comment: 'successful MongoDB injection' }
```

This attack will succeed in any case the username is correct, an assumption which is valid since harvesting user names isn't hard to achieve [15].

## VI. NoSQL JavaScript injection

A common feature of NoSQL databases is the ability to run javascript in the database engine in order to perform complicated queries or transactions such as map reduce. For example popular databases which allow this are MongoDB, CouchDB and its based descendants Cloudant [16] and BigCouch [17]. Javascript execution exposes a dangerous attack surface if un-escaped or not sufficiently escaped user input finds its way to the query. For example consider a complicated transaction which demanded javascript code and which includes an unescaped user input as a parameter in the query. As a use case let's take a model of a store which has a collection of items and each item has a price and an amount. The developer wanted to get the sum or average of these fields, so he writes a map reduce function that takes the field name that it should act upon (amount or price) as a parameter from the user. In PHP such code can look like this (where $param is user input):

```
$map = "function() {
  for (var i = 0; i < this.items.length; i++) {
      emit(this.name, this.items[i].$param); } }";
$reduce = "function(name, sum) { return Array.sum(sum); }";
$opt = "{ out: 'totals' }";
$db->execute("db.stores.mapReduce($map, $reduce, $opt);");
```

This code sums the field given by $param for each item by name. $param is expected to receive either "amount" or "price" for this code to behave as expected, but since user input is not being escaped here, a malicious input might include arbitrary javascript that will get executed. For Example, consider the following input:

```
a);}},function(kv) { return 1; }, { out: 'x'
});db.injection.insert({success:1});return
1;db.stores.mapReduce(function() { { emit(1,1
```

In its first part (in green) this payload closes the original map reduce function, then the attacker can execute any javascript he wishes on the database (in red) and eventually the last part (in blue) calls a new map reduce in order to balance the injected code into the original statement. After combining this user input into the string that gets executed we get (injected user input is colored in red):

```
db.stores.mapReduce(function() {
   for (var i = 0; i < this.items.length; i++) {
       emit(this.name, this.items[i].a);
   }
},function(kv) { return 1; }, { out: 'x' });
db.injection.insert({success:1});
return 1;db.stores.mapReduce(function() { { emit(1,1); } },
function(name, sum) { return Array.sum(sum); }, { out:
'totals' });"
```

This injection looks very similar to "classic" SQL injection. The defense against such an attack is disabling usage of javascript execution but if still required, properly escaping user input that finds its way into the code.

## VII. HTTP REST API AND ITS CONSEQUENCES

Another common feature of NoSQL databases is exposing an HTTP REST API that enables querying the database from client applications. For example, databases that expose a REST API include MongoDB, CouchDB and HBase. The exposure of a REST API enables simple exposure of the database to applications; even HTML5 only based applications, since it terminates the need for a mediate driver and allows any programming language to perform HTTP queries on the database. The advantages are clear, but does this feature come with a risk to security? We answer this on the affirmative: the REST API exposes the database to CSRF attacks allowing an attacker to bypass firewalls and other perimeter defenses. Let us examine how. As long as a database is deployed in a secure network behind security measures, such as firewalls, in order to compromise the database an attacker must either find a vulnerability that will let him into the secure network or perform an injection that will allow him to execute arbitrary queries. When a database exposes a REST API inside the secured network it allows anyone with access to the secured network to perform queries on the database using HTTP only – thus allowing such queries to be initiated even from the browser. If an attacker can inject an HTML form into a website or trick a user into a website of his own the attacker can perform any POST action on the database by submitting the form. POST actions include adding documents. For example, an attacker controls a malicious website and tricks an employee of company A to browse to that website, a technique called spear phishing (step 1 in Fig 2). Once the employee browses to the website a script submits an HTML form with an Action URL of an internal NoSQL DB (step 2). Since the employee is inside the secure network the DB is accessible for him and the action will succeed (step 3).

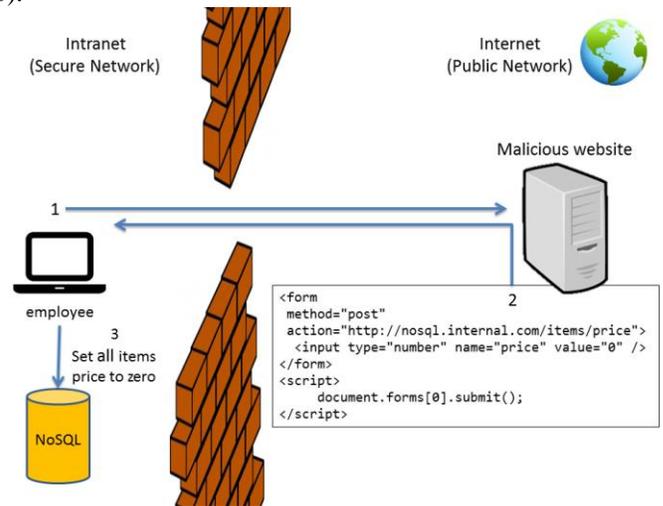

Figure 2: CSRF via NoSQL REST API

## VIII. MITIGATION

Mitigating security risks in NoSQL deployments is important in light of the attack vectors we presented in this paper. Let's examine a few recommendations for each of the threats:

### A. Security scanning to prevent injections

In order to mitigate injection attacks it is recommended to use out of the box encoding tools when building queries. For JSON queries such as in MongoDB and CouchDB almost all languages have good native encoding which will terminate the injection risk. It is also recommended to run Dynamic Application Security Testing (DAST) and static code analysis on the application in order to find any injection vulnerabilities if coding guidelines were not followed [18]. The problem is that many of the tools in the market today still lack rules for detecting NoSQL injections. DAST methodology is considered more reliable than static analysis [20], especially if used in conjunction with some backend inspection technology that improves detection reliability, a methodology referred to as Interactive Application Security Testing (IAST) [21][22].

### B. REST API exposure

To mitigate the risks of REST API exposure and CSRF attacks, there is a need to control the requests, limiting their format. For example, CouchDB has adopted some important security measures that mitigate the risk from having a REST API exposed. These measures include accepting only JSON in the content type. HTML forms are limited to URL encoded content type and hence an attacker will not be able to use html forms for CSRF and the other alternative is using AJAX requests and those are blocked by the browser thanks to same origin policy. It is also important to make sure JSONP and CORS are disabled in the server API to make sure that no actions can be made directly from a browser. It is important to note that some databases like MongoDB have many third party REST API's which are encouraged by the main project, some of these are really lacking in the security measures we described here.

### C. Access Control and Prevention of Privilege Escalation

In the past NoSQL did not support proper authentication and role management [9], today it is possible to manage proper authentication and RBAC authorization on most popular NoSQL databases. Utilizing these mechanisms is important for two reasons. First, they allow enforcing the principle of least privilege thus preventing privilege escalation attacks by legitimate users. Second, similarly to SQL injection attacks [19], proper privilege isolation allows to minimize the damage in case of data store exposure via the above described injections. Figure 3 illustrates an example in which the data accessible via a web application is authorized with a "user" role, while the sensitive entries require the "admin" role, which is never granted via the web interface. This allows scoping the damage in case of attack, ensuring that no administrators' data is leaked.

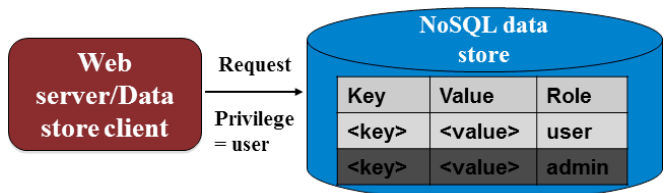

Figure 3: RBAC implemented on a NoSQL data store

### IX. SUMMARY

We have shown that NoSQL databases suffer from the same security risks as their SQL counterparts. Some of the low level techniques and protocols have changed but still the risks of injection, improper access control management and unsafe network exposure are high and similar between SQL and NoSQL systems. We recommend using mature databases with built-in security measures. However, even using the most secure data store does not prevent injection attacks which leverage vulnerabilities in the web applications accessing the data store. One way to prevent these is via careful code examination and static analysis. However, these may have high false positive rates and are difficult to conduct. While, dynamic analysis tools were shown to be very useful for the detection of SQL injection attacks [21], these should be adjusted to detect the specific vulnerabilities of NoSQL databases that we described in this paper.

REFERENCES

[1] "Imperva Web Application Attack Report" http://www.imperva.com/docs/HII_Web_Application_Attack_Report_Ed4.pdf
[2] MongoDB web site www.mongodb.org
[3] Redis Web site http://redis.io/
[4] Cassandra Web site http://cassandra.apache.org/
[5] Moniruzzaman, A. et. Al,. "Nosql database: New era of databases for big data analytics-classification, characteristics and comparison."
[6] Parker, Zachary, et. al. "Comparing nosql mongodb to an sql db." Proceedings of the 51st ACM Southeast Conference. ACM, 2013.
[7] No SQL and No Security https://www.securosis.com/blog/nosql-and-no-security
[8] Okman, Lior, et al. "Security issues in nosql databases." Trust, Security and Privacy in Computing and Communications (TrustCom), 2011 IEEE 10th International Conference on. IEEE, 2011.
[9] Factor, Michael, et al. "Secure Logical Isolation for Multi-tenancy in cloud storage." Mass Storage Systems and Technologies (MSST), 2013 IEEE 29th Symposium on. IEEE, 2013.
[10] DB-Engines Ranking, http://db-engines.com/en/ranking
[11] DB-Engines trends, http://db-engines.com/en/ranking_trend
[12] DB-Engines popularity changes, http://db-engines.com/en/ranking_categories
[13] MongoDB, Customers http://www.mongodb.com/industries/high-tech
[14] MongoDB, Sharding http://docs.mongodb.org/manual/sharding/
[15] "Invalid Username or Password": a useless security measure https://kev.inburke.com/kevin/invalid-username-or-password-useless/
[16] Cloudant web site https://cloudant.com/
[17] BigCouch web site http://bigcouch.cloudant.com/
[18] Static or Dynamic Application Security Testing? Both! http://blogs.gartner.com/neil_macdonald/2011/01/19/static-or-dynamic-application-security-testing-both/
[19] Least Privilege mitigation to SQL injection https://www.owasp.org/index.php/SQL_Injection_Prevention_Cheat_Sheet#Least_Privilege
[20] Haldar, Vivek, Deepak Chandra, and Michael Franz. "Dynamic taint propagation for Java." Computer Security Applications Conference, 21st Annual. IEEE, 2005.
[21] 9 Advantages of Interactive Application Security Testing (IAST) over Static (SAST) and Dynamic (DAST) Testing http://www1.contrastsecurity.com/blog/9-reasons-why-interactive-tools-are-better-than-static-or-dynamic-tools-regarding-application-security
[22] Glass Box: The Next Phase of Web Application Security Testing? http://www.esecurityplanet.com/network-security/glass-box-the-next-phase-of-web-application-security-testing.html
[23] MongoDB documentation on Security http://docs.mongodb.org/manual/core/security-introduction/
[24] J. H. Saltzer and M. D. Schroeder. The protection of information in computer systems. In Proc. IEEE, 63(9):1278–1308, 1975.